\documentclass[%
 reprint,
 amsmath,amssymb,
 aps,
]{revtex4}

\usepackage[normalem]{ulem}

\usepackage{graphicx}
\usepackage{dcolumn}
\usepackage{bm}
\usepackage{color}
\usepackage[dvipsnames]{xcolor}
\usepackage{hyperref}
\hypersetup{
  colorlinks=true,        
   linkcolor=ForestGreen,    
}

\usepackage{multirow,makecell}
\newcommand{\onbb}{0$\nu\beta\beta$}

\begin{document}

\preprint{APS/123-QED}

\title{Renormalizability of the leading order operator for neutrinoless double beta decay with the effects of finite nucleon size}

\author{Tai-Xing Liu$^a$, Ri-Guang Huang$^{a,b}$ and Dong-Liang Fang$^{a,b}$}
\email{dlfang@impcas.ac.cn}

\address{$^a$Institute of Modern Physics, Chinese Academy of Sciences, Lanzhou 730000, China}
\address{$^b$School of Nuclear Sciences and Technology, University of Chinese Academy of Sciences, Beijing 101408, China}



\date{\today}

\begin{abstract}

The fundamental process of neutrinoless double beta decay, $nn\to ppe^-e^-$, dominated by the exchange of light Majorana neutrinos, is studied in the framework of chiral effective field theory. Considering neutrinos as virtual states, we evaluate the contributions of finite nucleon size to the transition amplitude in a non-perturbative manner,
as opposed to expanding these effects in powers of momentum. Based on the nucleon form factors expressed in terms of the dipole and Kelly parametrizations, we find that, at the leading order, the present scheme could renormalize the amplitude in the context of the standard mechanism and provide predictions consistent with the previous investigations. Consequently, we argue that the impact of the effects of finite nucleon size on the amplitude is comparable to that of the leading-order contact term introduced in 
\cite{Cirigliano:2018hja}
in a perturbative scheme. Our results provide not only a benchmark calculation for the transition amplitude between two schemes but also evidence for the reasonableness of non-perturbative treatment for the effects of finite nucleon size in conventional nuclear many-body methods.

\end{abstract}

\maketitle

\emph{Introduction.}\textemdash Neutrinoless double beta ({\onbb}) decay, a hypothetical second-order weak transition that converts two neutrons in a nucleus into two protons through the emission of two electrons but without neutrinos~\cite{Furry:1939qr}, is an ideal venue for probing lepton number violation (LNV). Besides demonstrating that neutrinos are Majorana fermions~\cite{Schechter:1981bd}, its observation provides insightful perspectives to explain the neutrino mass generation~\cite{Mohapatra:1979ia}, determines the absolute neutrino mass scale and hierarchy~\cite{Avignone:2007fu}, and explores the mechanism of the matter-antimatter asymmetry in early universe~\cite{Davidson:2008bu}. Currently, numerous underground experiments with high sensitivity~\cite{PandaX-II:2019euf, GERDA:2020xhi, CUORE:2020ymk, KamLAND-Zen:2022tow} all over the world are actively searching for {\onbb} decay signals for different even-even nuclei. In the future, more ton-scale facilities~\cite{LEGEND:2017cdu, nEXO:2017nam, CUPID:2022wpt, NnDEx-100:2023alw} will be deployed to improve the discovery potential of experiments and measure the half-life of different candidate isotopes (for a recent review, see Ref.~\cite{Agostini:2022zub}).

These experiments will help us make better understanding of this process. In general, whatever operator including the LNV physics may induce the nuclear {\onbb} decay. To reveal the underlying mechanism of this process, the precise calculation of nuclear matrix elements (NMEs) that incorporates various LNV contributions is essential. However, its estimation based on the different many-body methods suffers from a discrepancy of a factor of three to five thus far~\cite{Engel:2016xgb}. One should note that the uncertainty of NMEs is rooted in the realization of nuclear structure and LNV mechanisms.

It has been shown that the chiral effective field theory (ChEFT)~\cite{Weinberg:1990rz, Weinberg:1991um} offers a promising tool to quantify such uncertainties in nuclear calculations, for a comprehensive review see Ref.~\cite{Hammer:2019poc}. Moreover, the electroweak currents can be introduced systematically in ChEFT~\cite{Pastore:2009is, Baroni:2015uza, Krebs:2016rqz, Krebs:2019aka}. By using series of EFTs, the various LNV sources at energy well above the electroweak scale can be expressed as the {\onbb} operators at the nucleon level~\cite{Prezeau:2003xn, Cirigliano:2017djv, Cirigliano:2017tvr, Cirigliano:2018yza, Dekens:2020ttz, deVries:2022nyh}. With ChEFT, \textit{ab initio} calculations under the standard light Majorana neutrino-exchange mechanism~\cite{Weinberg:1979sa} for NMEs were performed~\cite{Yao:2019rck, Belley:2020ejd, Novario:2020dmr, Wirth:2021pij, Belley:2023lec}.

Recently, within the standard mechanism of {\onbb} decay, the ChEFT description based on naive dimensional analysis (NDA)~\cite{Cirigliano:2017tvr} has been proven to violate the requirement of renormalizability~\cite{Cirigliano:2018hja, Cirigliano:2019vdj}. As a consequence, a contact operator that encodes the effect of hard-neutrino exchange is promoted to leading order (LO). The corresponding low-energy constant (LEC) is currently still undetermined by experiments or lattice QCD simulations, introducing additional uncertainties to the estimation of NMEs besides those from nuclear structure.

The prescription to estimate the size of the LEC was first presented by Cirigliano \textit{et al.} in~\cite{Cirigliano:2020dmx, Cirigliano:2021qko}. Although this approach by matching the $nn\to ppe^-e^-$ amplitude onto the Cottingham formula~\cite{Cottingham:1963zz}, is somewhat model-dependent, the uncertainty is acceptable at LO calculation for ChEFT. On the other hand, Yang \textit{et al.}~\cite{Yang:2023ynp} developed a relativistic method that can renormalize the LO decay amplitude by resorting to the light neutrino exchange rather than utilizing a contact term. The two distinct approaches show decent consistency in the prediction of the LO amplitude.

An essential ingredient used to evaluate the {\onbb} amplitude is the coupling of lepton-neutrino current to a nucleon weak current. This coupling can be decomposed into all structures permitted by Lorentz invariance, whose kinematics is factorized into the momentum-dependent functions, i.e., nucleon form factors (FFs), encoding the effects of finite nucleon size. As an application of ChEFT for the low-energy electroweak processes for nucleons, the nucleon FFs can be expressed in terms of powers of momentum (see Ref.~\cite{Bernard:1995dp} for a review), in which the LO term corresponds to the point-like assumption of the nucleon. It is important to emphasize that the prerequisite for this perturbative expansion is the observability of the external source particles. If one of the particles is virtual, the nucleon FFs will vary based on the momentum of the intermediate state. From this perspective, a non-perturbative approach to incorporate nucleon structure effects may provide novel insights into the renormalization of the {\onbb} amplitude. 
{Furthermore, it is worth stressing that the aforementioned nucleon FFs are inappropriate for assessing the impact of nucleon structure on nuclear forces because nuclear forces are so strong that the internal structure of the nucleon changes during strong process.}

In this study, we revisit the $nn\to ppe^-e^-$ transition mediated solely by the light Majorana neutrinos to examine the renormalizability of {\onbb} operators. The finite size of the nucleon is taken into account in a non-perturbative manner, similar to methodologies employed in nuclear many-body calculations~\cite{Simkovic:1999re, Simkovic:2009pp, Hyvarinen:2015bda, Yao:2021wst}. Using the momentum-dependent FFs, we obtain the finite and regulator-independent amplitude at LO in the absence of short-range contribution. It shows the model-free advantage of this method and the validity of the way to treat nucleon structure in conventional nuclear calculation.

\emph{Theoretical framework.}\textemdash We focus on the scenario of light Majorana neutrino mediated $nn\to ppe^-e^-$ transition. The scattering of initial and final state particles is restricted to the $^1S_0$ channel because it has been proved in Refs.~\cite{Cirigliano:2018hja, Cirigliano:2019vdj} that the conflict between NDA and renormalizability only appears in this channel. For this purpose, the heavy-baryon formulation~\cite{Jenkins:1990jv, Bernard:1992qa} of ChEFT is employed and the Coulomb interaction is ignored.

We start with the chiral Lagrangian, the nucleon-nucleon potential derived from two-nucleon irreducible diagrams comprises the long-range pion exchange and the short-range four-nucleon contact terms. At LO, the two-nucleon potential is given by
\begin{equation}
    \begin{split}
        V_{NN}(\pmb{p}',\pmb{p})=&-\frac{g_A^2}{4f_\pi^2}\vec{\tau}_1\cdot\vec{\tau}_2\frac{\pmb{\sigma}_1\cdot\pmb{q}\,\pmb{\sigma}_2\cdot\pmb{q}}{\pmb{q}^2+m_\pi^2}\\
        &+C_S+C_T\,\pmb{\sigma}_1\cdot\pmb{\sigma}_2\,,
    \end{split}
    \label{LOnnpot}
\end{equation}
where $g_A=1.27$ is the nucleon axial coupling, $f_\pi=92.2$ MeV the pion decay constant, $m_\pi=138$ MeV the pion mass. The initial and final relative momenta, denoting as $\pmb{p}$ and $\pmb{p}\,'$ respectively, define the momentum transfer $\pmb{q}=\pmb{p}\,'-\pmb{p}$. In the $^1S_0$ channel, the combination of LECs $C_S$ and $C_T$, i.e. $\widetilde{C}=C_S-3C_T$, is needed.

In the context of the standard mechanism of {\onbb} decay, the electron-neutrino Majorana mass term, which is the dominant contribution of LNV at low energy, can be written as
\begin{equation}
    \mathcal{L}_{\Delta L=2}=-\frac{m_{\beta\beta}}{2}\nu_{eL}^TC\nu_{eL}\,,
    \label{numass}
\end{equation}
where $C=i\gamma_2\gamma_0$ denotes the charge conjugation operator, and the effective mass $m_{\beta\beta}=\sum_iU^2_{ei}m_i$ is defined with masses of three generations of light neutrino and elements of the neutrino mixing matrix. The definition of the neutrino potential is similar to that of the two-nucleon potential~\cite{Cirigliano:2017djv}. At LO in NDA, only following neutrino-exchange potential
\begin{equation}
    \begin{split}
        V_\nu(\pmb{p}',\pmb{p})=&\tau_1^+\tau_2^+\frac{1}{\pmb{q}^2}\left[G_V^2\left(\pmb{q}^2\right)-G_A^2\left(\pmb{q}^2\right)\,\pmb{\sigma}_1\cdot\pmb{\sigma}_2\right.\\
        &\left.+G_A^2\left(\pmb{q}^2\right)\,\pmb{\sigma}_1\cdot\pmb{q}\,\pmb{\sigma}_2\cdot\pmb{q}\frac{2m_\pi^2+\pmb{q}^2}{\left(\pmb{q}^2+m_\pi^2\right)^2}\right]
    \end{split}
    \label{LOnupot}
\end{equation}
contributes to the $nn\to ppe^-e^-$ amplitude. In light of the non-perturbative property of the nucleon structure in the {\onbb} process, we have introduced the momentum-dependent FFs $G_V\left(\pmb{q}^2\right)$ and $G_A\left(\pmb{q}^2\right)$, which are normalized as the vector and axial-vector coupling respectively, to take into account the finite-nucleon-size effects.

The commonly used dipole parametrization $G_\alpha\left(\pmb{q}^2\right)=g_\alpha/(1+\pmb{q}^2/\Lambda_\alpha^2)^2$ ($\alpha=V,A$) is employed in this work. Here we adopt $\Lambda_V=0.84$ GeV and $\Lambda_A=1.05$ GeV respectively. Given that the prediction of dipole parametrization is inconsistent with data in the high-momentum region~\cite{JeffersonLabHallA:1999epl, Gayou:2001qt, JeffersonLabHallA:2001qqe, Puckett:2010ac, Meyer:2016oeg}, we additionally utilize the parametrization scheme presented by Kelly~\cite{Kelly:2004hm} to describe the structure of the nucleon. 

The Kelly parametrization of the axial FF takes the form
\begin{equation}
    G_A(x)=\frac{g_A+a_1x}{1+b_1x+b_2x^2+b_3x^3}\,,
\end{equation}
where $x=\pmb{q}^2/(4m_N^2)$ with the nucleon mass $m_N=939$ MeV, the dimensionless model parameters $a_1,b_1,b_2$ and $b_3$ are determined by fitting of data from Ref.~\cite{Hashamipour:2019pgy}. On the other hand, in conjunction with the fits in Refs.~\cite{Kelly:2004hm, Punjabi:2015bba} for the Sachs FFs of proton and neutron, one can obtain the vector FF as a function of momentum via the relation
\begin{equation}
    G_V(x)=\frac{\left[G_{Ep}(x)-G_{En}(x)\right]+x\left[G_{Mp}(x)-G_{Mn}(x)\right]}{1+x}\,,
\end{equation} 
where $G_{Ep}$, $G_{Mp}$, $G_{En}$ and $G_{Mn}$ are the so-called electric and magnetic Sachs FFs, for proton and neutron, respectively. 

\begin{figure}[htbp]
\centering
\begin{minipage}[t]{0.48\textwidth}
\centering
\includegraphics[scale=0.37]{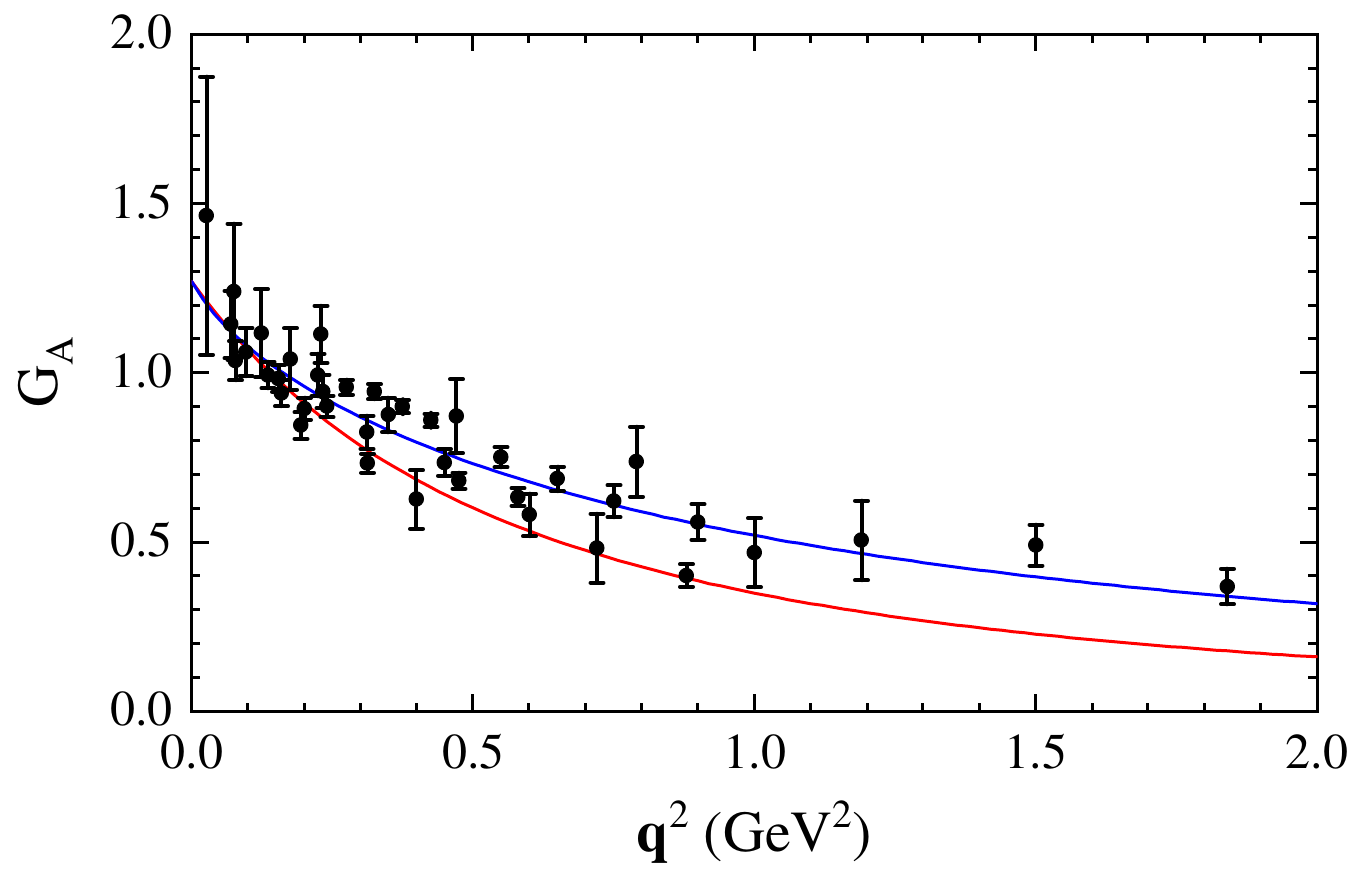}
\end{minipage}\\
\vspace{10pt}
\begin{minipage}[t]{0.48\textwidth}
\centering
\includegraphics[scale=0.37]{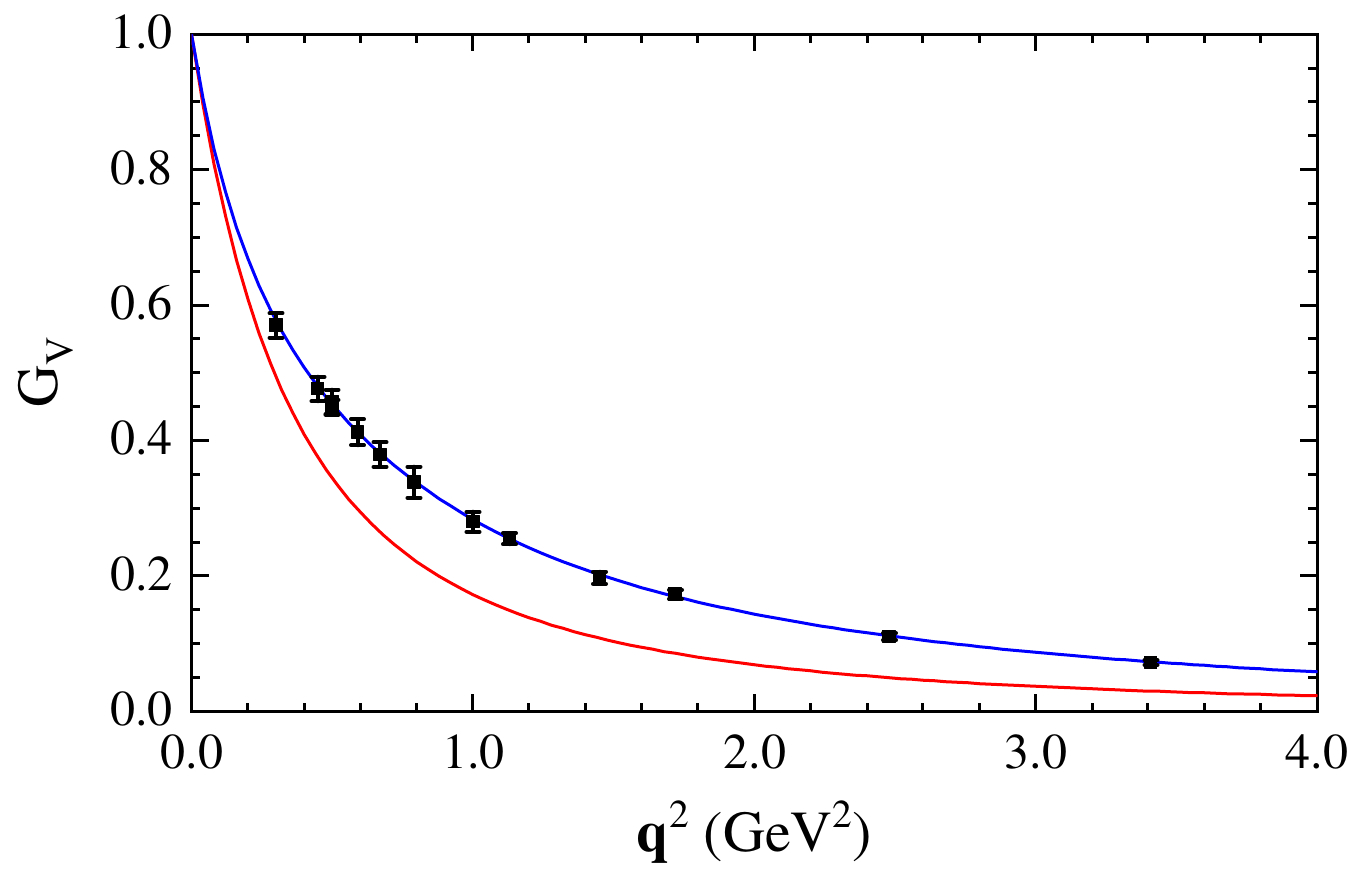}
\end{minipage}
\caption{Dependence of the nucleon axial (top) and vector (bottom) FFs on momentum. The solid circles and squares with error bars represent the experimental data of $G_A$ and $G_V$, respectively, from Ref.~\cite{Hashamipour:2019pgy} and Ref.~\cite{Cates:2011pz}. The red and blue lines represent the predictions of the dipole and Kelly parametrizations, respectively.}
\label{fig:nffs}
\end{figure}

The results of the dipole and fitted Kelly forms are shown in Fig.\ref{fig:nffs}. 
The prediction of the Kelly parametrization for the nucleon FFs is in reasonable agreement with the data. One of the important points is that both types of parametrization schemes for vector and axial FFs all predict a general behavior: 
\begin{equation}
    G_\alpha\propto\pmb{q}^{-4}\,,\,\,\,\,\text{for}\,\,\vert \pmb{q}\vert\to\infty\,.
\end{equation} 
By this scale rule of FFs, we will demonstrate the convergence of the {\onbb} amplitude in the ultraviolet (UV) region. Comparing the transition amplitudes based on both parametrizations, we will demonstrate the validity of assessing the effects of finite nucleon size with dipole parametrization on NMEs. 

The LO {\onbb} amplitude based on the kinematics $n(\pmb{p})n(-\pmb{p})\to p(\pmb{p}')p(-\pmb{p}')e^-\left(\pmb{p}_{e1}=0\right)e^-\left(\pmb{p}_{e2}=0\right)$ can be schematically expressed as in \cite{Cirigliano:2019vdj}:
\begin{equation}
    A_\nu=-\left(V_\nu+V_\nu G_0T_s+T_sG_0V_\nu+T_sG_0V_\nu G_0T_s\right)\,,
    \label{anuamp}
\end{equation}
where $G_0$ is the two-nucleon free propagator, $T_s$ the two-nucleon scattering amplitude obtained by iterating the LO strong potential $V_{NN}$ through $G_0$. We name the four terms in Eq.\eqref{anuamp}, from left to right, $A_\mathcal{A}$, $A_\mathcal{B}$, $\bar{A}_\mathcal{B}$, and $A_\mathcal{C}$.

We are interested in the renormalizability of the {\onbb} amplitude. From Eq.\eqref{anuamp}, the sub-amplitude $A_\mathcal{A}$ as a tree-level contribution must be convergent. Because of the finiteness of the strong amplitude $T_s$, the UV behavior of the remaining terms is dominated by the loop integrals with the insertion of neutrino potential. The nonrelativistic propagator $G_0$ behaves like $\mathcal{O}\left(\Lambda^{-2}\right)$ in the UV region, where $\Lambda$ denotes a UV energy scale much greater than the typical momentum in the process. For the neutrino potential in the $^1S_0$ wave, its UV scale rule is given by
\begin{equation}
     V_\nu\sim\mathcal{O}\left(\Lambda^{-2-2g}\right)\,,
\end{equation}
where $g=0$ (4) corresponds to the case without (with) the effects of finite nucleon size. Therefore, in the nonrelativistic case, the superficial degree of divergence for the last three terms in Eq.\eqref{anuamp} has the following form:
\begin{equation}
    D=L-2N(1+g)\,,
\end{equation}
where $L$ is the number of loops, and $N=1$ is the number of insertions of the neutrino potential. Given that $L=1$ and $g\geq0$ for $A_\mathcal{B}$ and $\bar{A}_\mathcal{B}$, it follows that the both sub-amplitudes could reach the convergence as $\mathcal{O}\left(\Lambda^{-1-2g}\right)$. However, $A_\mathcal{C}\sim\mathcal{O}\left(\Lambda^{-2g}\right)$ means that its convergence depends on the treatment of the nucleon structure. In the case without the effects of finite nucleon size, i.e. $g=0$, $A_\mathcal{C}$ diverges logarithmically such that a LEC $g_\nu^{NN}\sim\mathcal{O}\left(f_\pi^{-2}\right)$ is required to renormalize the LO amplitude $A_\nu$~\cite{Cirigliano:2018hja, Cirigliano:2019vdj}. With finite nucleon size, however, $A_\mathcal{C}\sim\mathcal{O}\left(\Lambda^{-8}\right)$ is finite so that $A_\nu$ can be renormalized without inclusion of any short-range counter terms. 

\emph{Results and discussions.}\textemdash Following the usual numerical strategy, the renormalizability of $A_\nu$ is demonstrated by its independence of the cutoff value in momentum or coordinate space. In this work, we introduce the momentum cutoff $\Lambda$ via the way that regulates the UV part of the LO strong potential with a separable Gaussian regulator:
\begin{equation}
    V_{NN}(\pmb{p}',\pmb{p})\longrightarrow \text{exp}\left(-\frac{\pmb{p}'\,^4}{\Lambda^4}\right)V_{NN}(\pmb{p}',\pmb{p})\,\text{exp}\left(-\frac{\pmb{p}^4}{\Lambda^4}\right).
\end{equation}
To determine the LEC $\widetilde{C}$ in the strong interaction, we fit $T_s$ to the $^1S_0$ wave neutron-proton scattering length $a_{np}=-23.74$ fm. The cutoff independence of the phase shift has been achieved at $\Lambda\to\infty$.

\begin{figure}[htbp]
    \centering
    \includegraphics[scale=0.37]{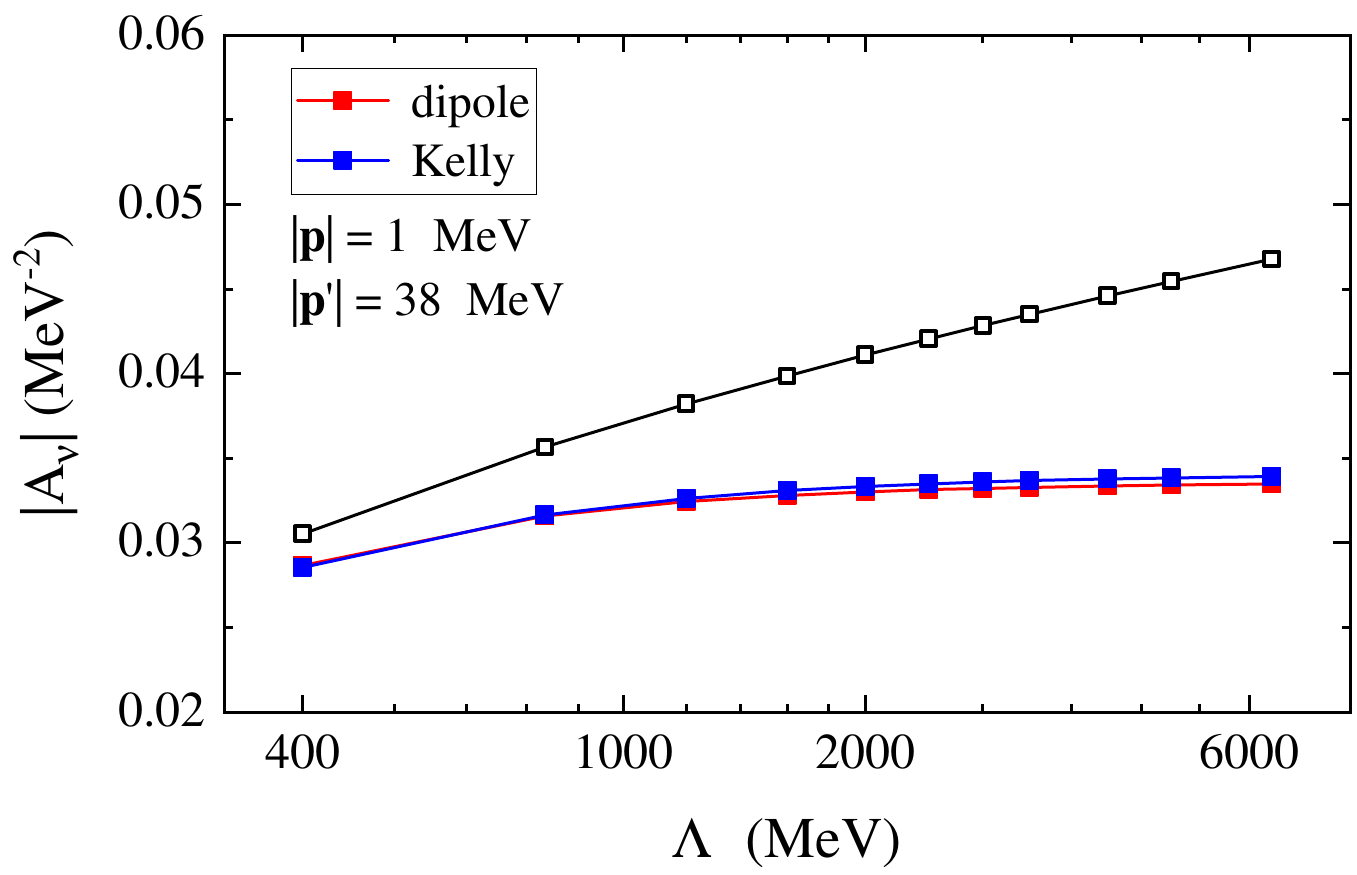}
    \caption{The LO amplitude $A_\nu$ as a function of the cutoff value $\Lambda$. The open squares are the results without the contribution of finite nucleon size. The results denoted by red and blue solid squares take into account the effects of finite nucleon size, via the dipole and Kelly parametrizations, respectively.}
    \label{fig:amp138}
\end{figure}

Choosing the kinematic points $\vert\pmb{p}\vert=1$ MeV and $\vert\pmb{p}'\vert=38$ MeV, we depict in Fig.\ref{fig:amp138} the cutoff variation of the LO amplitude $A_\nu$.
The ChEFT results without the contributions of nucleon structure display a logarithmic divergence, while the amplitude with the finite-nucleon-size effects shows a satisfactory convergence, as same as the expectation above. In addition, the amplitudes obtained from both the dipole and Kelly parametrizations are remarkably similar at any cutoff value, with a difference around only 1\%. This indicates that the impact of different parametrizations of nucleon structure on {\onbb} decay is negligible, highlighting the robustness and reliability of the parametrization in calculating the {\onbb} amplitude. Therefore, the dipole parametrization is accurate enough to assess the contribution of the nucleon structure to NMEs.

\begin{figure}[htbp]
    \centering
    \includegraphics[scale=0.37]{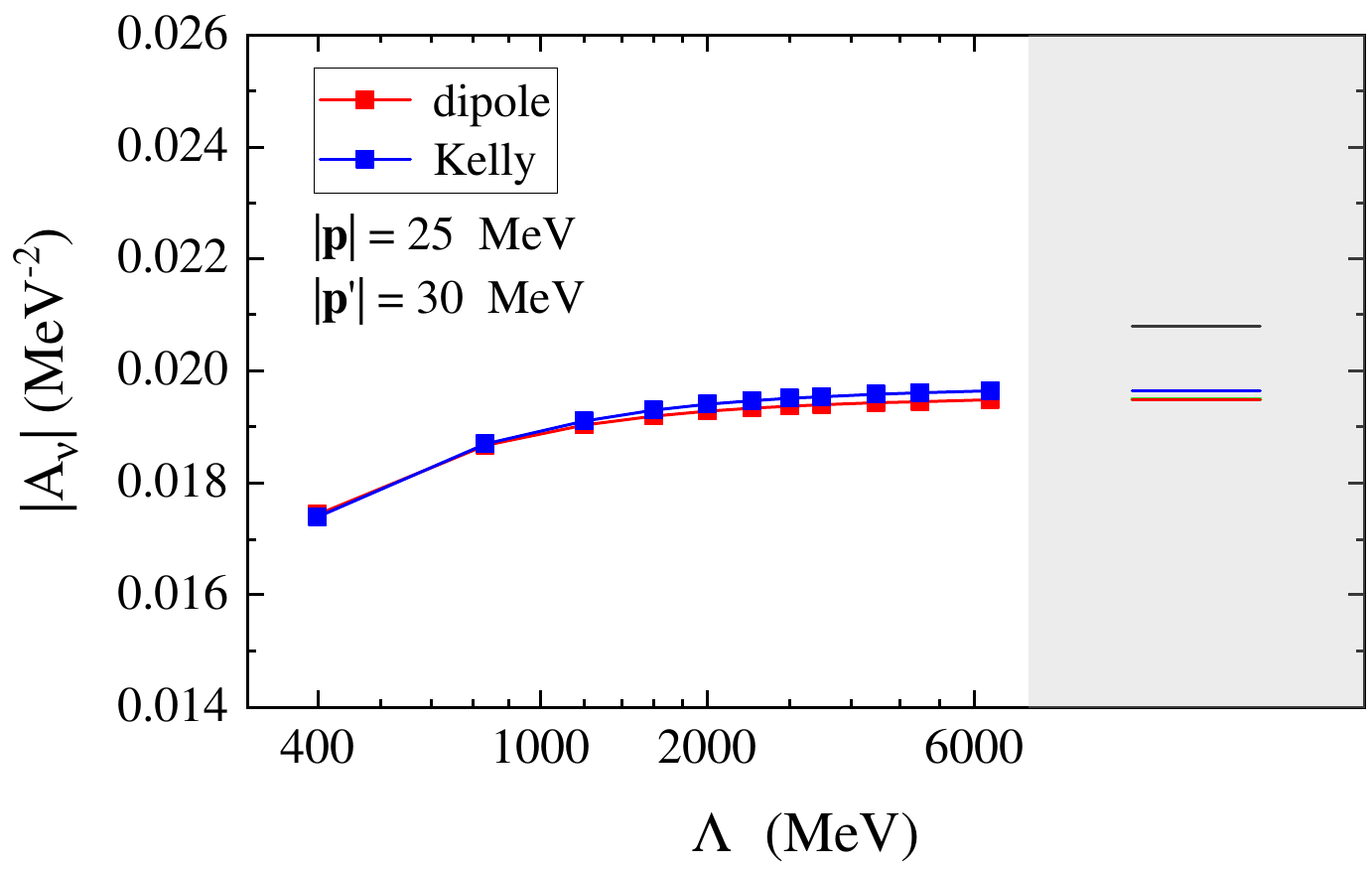}
    \caption{The LO amplitude $A_\nu$ as a function of the cutoff value $\Lambda$, in comparison with the results taken from Ref.~\cite{Cirigliano:2020dmx} (green line) and Ref.~\cite{Yang:2023ynp} (black line), respectively.
    }
    \label{fig:amp2530}
\end{figure}

To validate the present scheme, we calculate the amplitude for $\vert\pmb{p}\vert=25$ MeV and $\vert\pmb{p}'\vert=30$ MeV and perform a comparison with the previous results~\cite{Cirigliano:2020dmx, Cirigliano:2021qko, Yang:2023ynp}. As shown in Fig.\ref{fig:amp2530},
again, we capture the renormalized amplitude without introducing any undeterminate contact operators. The limit value of absolute amplitude $\vert A_\nu\vert=0.01956(8)$ $\text{MeV}^{-2}$ at this kinematic point, where the error stems from the dependence on parametrization. Although the approach used to renormalize $A_\nu$ in this study differs from that in Refs.~\cite{Cirigliano:2020dmx, Yang:2023ynp}, our results are consistent with the estimations of Refs.~\cite{Cirigliano:2020dmx, Yang:2023ynp} within the uncertainty 10\%, a relatively strict upper bound for ChEFT truncation error at LO. 

\begin{figure}[htbp]
\centering
\begin{minipage}[t]{0.48\textwidth}
\centering
\includegraphics[scale=0.37]{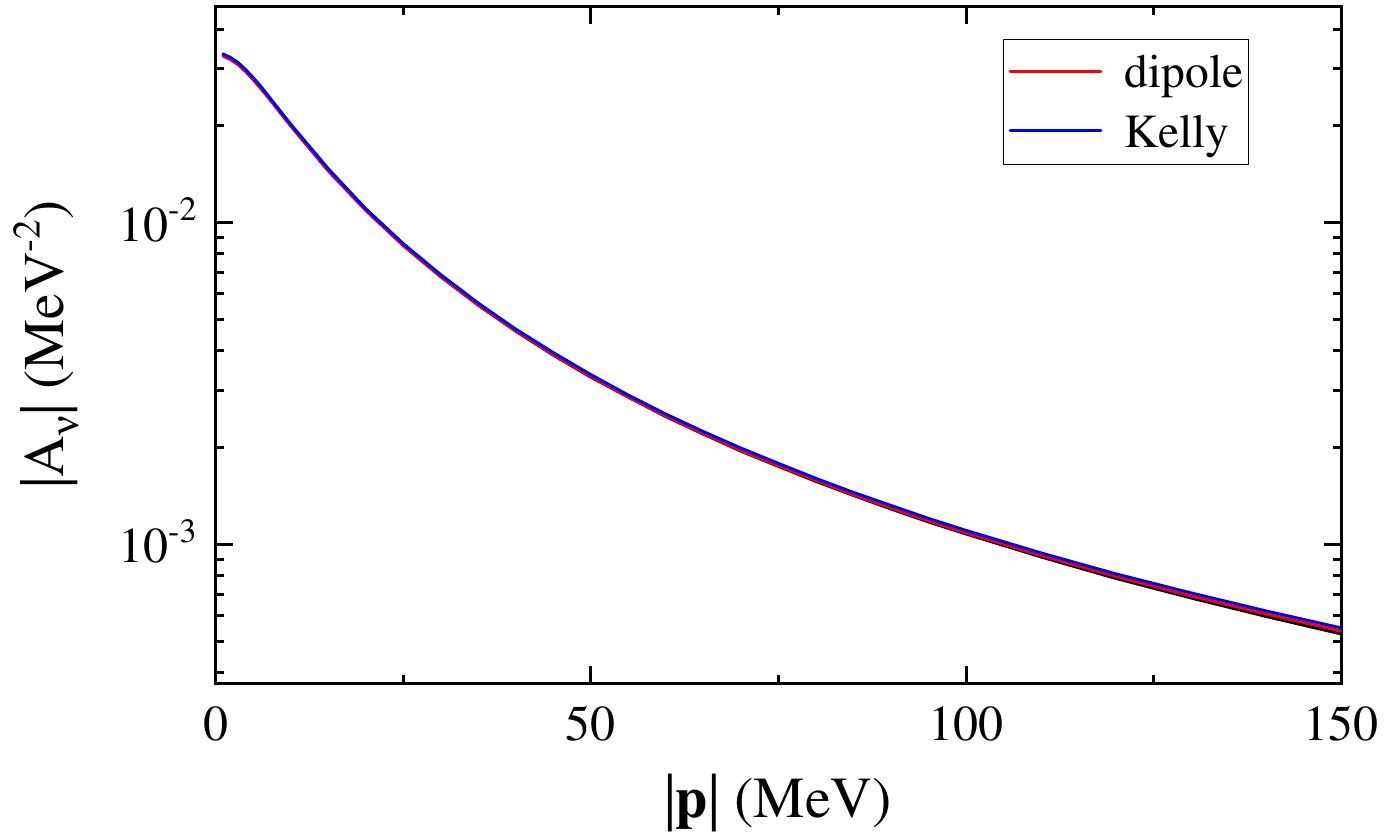}
\end{minipage}\\
\vspace{10pt}
\begin{minipage}[t]{0.48\textwidth}
\centering
\includegraphics[scale=0.37]{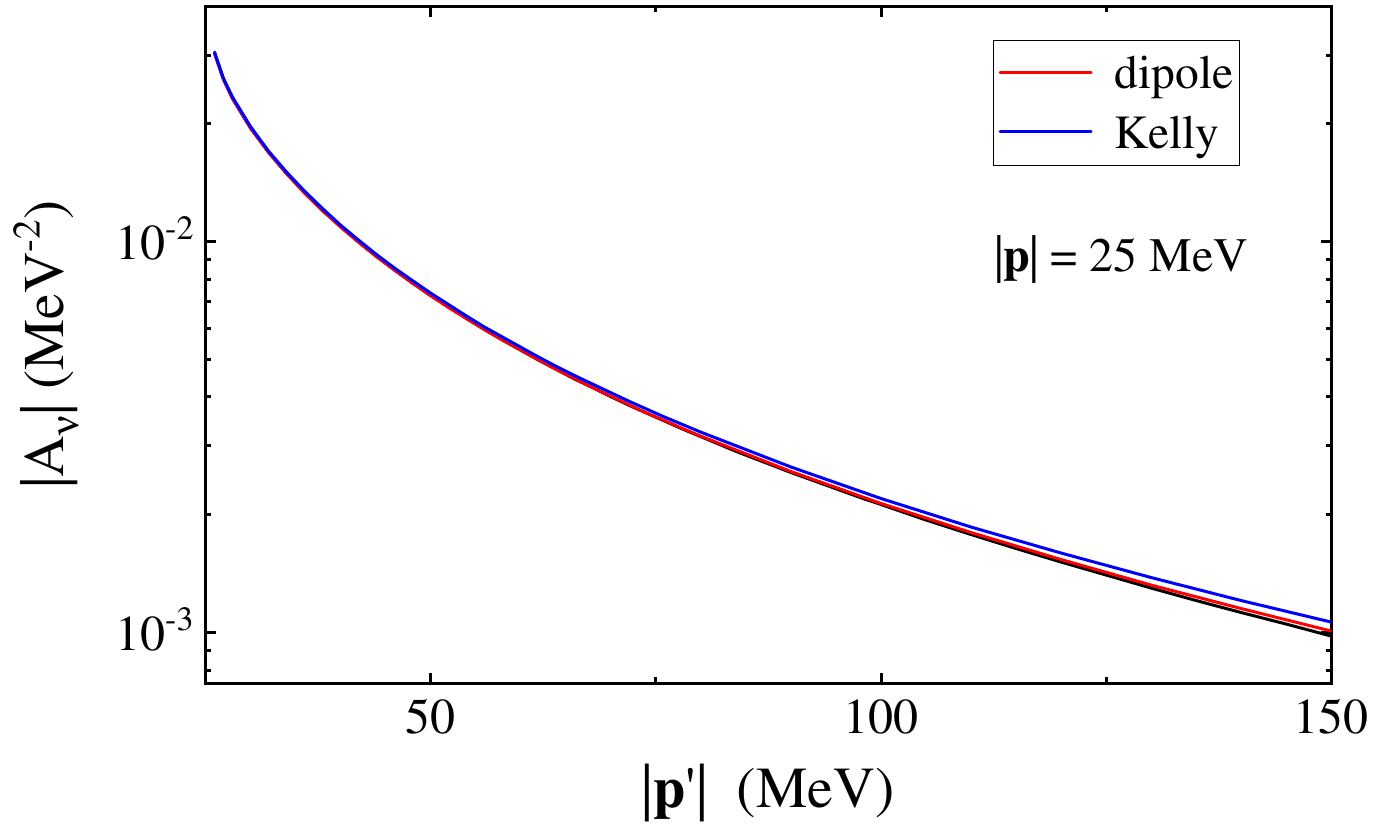}
\end{minipage}
\caption{Dependence of the on-shell (top panel) and off-shell (bottom panel) amplitudes on momentum for cutoff value $\Lambda=6.4$ GeV. The black line denotes the results based on the long- and short-range transition operators, in which the LEC is determined by reproducing the amplitude at $\vert\pmb{p}\vert=25$ MeV and $\vert\pmb{p}'\vert=$ 30 MeV, i.e., $A_\nu$ = -0.0195 $\text{MeV}^{-2}$~\cite{Cirigliano:2020dmx}.
}
\label{amp_anu}
\end{figure}

The renormalized on-shell and off-shell amplitudes as functions of momentum are shown in Fig.\ref{amp_anu}. At various kinematic points, the results accounting for the effects of nucleon structure are in quite good agreement with the predictions based on the renormalized scheme of Refs.~\cite{Cirigliano:2020dmx, Cirigliano:2021qko}. 
Furthermore, the present scheme can also recover the transition amplitude for higher partial-wave channels obtained in Ref~\cite{Cirigliano:2019vdj}.

The primary motivation for employing a non-perturbative approach to treat the finite-size effects of the nucleon arises from the inadequacy of perturbative expansion of the nucleon FFs caused by virtual neutrinos, rather than from the behavior of FFs decreasing as $\mathcal{O}(\Lambda^{-4})$ in the UV region. The numerical results indicate that the approach to including the suppressed effects of FFs is equivalent to the renormalized scheme of Ref.~\cite{Cirigliano:2020dmx} at LO. Therefore, to ensure the correctness of the result, we argue that the calculation of the {\onbb} amplitude should be conducted within a uniform renormalization scheme.

We note that conventional many-body calculations estimate the contributions of finite nucleon size to NMEs in the same manner as this work. Based on the above discussions, we conclude that these calculations introducing the nucleon structure effects in a non-perturbative way are consistent with the scheme that renormalizes the amplitude with short-range terms. In contrast, combining the nucleon structure with the short-range contribution in many-body calculation may lead to an overestimation of the amplitude, thereby amplifying the uncertainty of NMEs.

Moreover, since we evaluate the LO {\onbb} amplitude without relying on any unknown LECs, our approach provides a model-independent description for {\onbb} decay. This is crucial for improving the precision of NMEs calculated in many-body methods. Besides circumventing model dependence, the realistic size of the short-range operator as discussed in Refs.~\cite{Cirigliano:2018hja, Cirigliano:2019vdj, Cirigliano:2020dmx, Cirigliano:2021qko} can be addressed by matching the estimation of the decay rate based on the present scheme.

Our framework is more straightforward for advancing higher-order calculations, in contrast to the relativistic scheme~\cite{Yang:2023ynp} within the framework of manifestly Lorentz-invariant ChEFT. The latter would introduce certain complexities when moving to higher orders, especially for the treatment of the renormalization~\cite{Hammer:2019poc, Epelbaum:2019kcf}. 
Meanwhile, our method facilitates a more intuitive combination with existing many-body techniques compared to the relativistic scheme.

\emph{Summary.}\textemdash In this study, we present a nonrelativistic approach to renormalize the LO {\onbb} amplitude for the standard mechanism using the effects of nucleon structure instead of the short-distance operators with undetermined LECs. Our calculations of the {\onbb} amplitude at various kinematic points demonstrate the efficacy of the proposed scheme, showing its comparability with established frameworks~\cite{Cirigliano:2020dmx, Cirigliano:2021qko}. The most significant implication of this equivalence is the validation of conventional many-body calculations based on the light Majorana neutrino mechanism, 
confirming the absence of any oversight regarding LO LNV physics in these calculations. Finally, our scheme has the capacity to transcend the specific scenario of light Majorana neutrino exchanges, generalizing to the non-standard decay mechanisms and helping to reduce the uncertainties of corresponding NMEs in many-body calculations.

\begin{acknowledgments}
We thank Prof. F. \v{S}imkovic for inspiring us to perform this calculation. This work is supported by the National Key Research and Development Program of China (2021YFA1601300), the Chinese Academy of Sciences with Project for Young Scientists in Basic Research (YSBR- 099), and the Postdoctoral Special Foundation of Gansu Province.
\end{acknowledgments}

\bibliography{0nbbLO.bib}

\end{document}